\documentclass[final,3p,times]{elsarticle}
\usepackage{graphicx}
\usepackage{amssymb}
\usepackage{amsmath}
\usepackage{hyperref}
\usepackage{color}

\journal{Annals of Physics (accepted)}

\begin{document}

\begin{frontmatter}

        %%%%%%%%%%%%%%%%%%%%%%%%%%%%%%%%%%%%%%%%%%%%%%%%%%%%%%%%%%%%%%%%%%%%%%%%%%%%%%
        %%%%                     Title and authors                                %%%%
        %%%%%%%%%%%%%%%%%%%%%%%%%%%%%%%%%%%%%%%%%%%%%%%%%%%%%%%%%%%%%%%%%%%%%%%%%%%%%%

        \title{Limitation of the Lee-Huang-Yang interaction in forming a self-bound state in     Bose-Einstein condensates}

        \author[san]{Sandeep Gautam}
\ead{sandeep@iitrpr.com}
   
\author[ift]{Sadhan K. Adhikari\corref{author}}
\ead{sk.adhikari@unesp.br}
\cortext[author]{Corresponding author}

     \address[san]{Department of Physics, Indian Institute of Technology Ropar, Rupnagar, Punjab 140001,
                     India              }
     \address[ift]{IFnstituto de F\'{\i}sica Te\'orica, Universidade Estadual
                     Paulista - UNESP,  01.140-070 S\~ao Paulo, S\~ao Paulo, Brazil}

     %   \author{S. K. Adhikari\footnote{adhikari44@yahoo.com,
      %            http://www.ift.unesp.br/users/adhikari}}

        %%%%%%%%%%%%%%%%%%%%%%%%%%%%%%%%%%%%%%%%%%%%%%%%%%%%%%%%%%%%%%%%%%%%%%%%%%%%%%
        %%%%%%%%%%                    Abstract                             %%%%%%%%%%%
        %%%%%%%%%%%%%%%%%%%%%%%%%%%%%%%%%%%%%%%%%%%%%%%%%%%%%%%%%%%%%%%%%%%%%%%%%%%%%%

        \date{\today}
     \begin{abstract}

The perturbative  Lee-Huang-Yang (LHY) interaction   proportional to $n^{3/2}$, where $n$ is the density, creates an infinitely repulsive potential at the center of a  Bose-Einstein condensate (BEC) with net attraction, which stops the collapse to form a  self-bound state  
  in a dipolar BEC and in a binary BEC. 
 However,  recent microscopic  calculations of  the non-perturbative  beyond-mean-field (BMF) interaction indicate that the LHY interaction %with the $n^{3/2}$ term 
is valid only for very  small values of gas parameter $x$.   We show that a realistic non-perturbative BMF interaction can stop collapse and   form a  self-bound state only in a weakly attractive binary BEC with small $x$ values  ($x\lessapprox 0.01$), whereas the perturbative  LHY interaction stops collapse for all attractions.  
We demonstrate 
these aspects using an analytic  BMF interaction   with appropriate  weak-coupling LHY and strong coupling limits.

\end{abstract}

%\pacs{03.75.Mn, 03.75.Hh, 67.85.Bc, 67.85.Fg}

\begin{keyword}

Self-bound binary Bose-Einstein condensate, Beyond-mean-field interaction,
Lee-Huang-Yang interaction, Collapse instability

\end{keyword}

\end{frontmatter}

  %   \maketitle
% \section{Introduction}
%\label{Sec-I}

{\section {Introduction}}
A one-dimensional (1D) bright soliton, formed  due to a balance 
between defocusing forces and 
 nonlinear attraction, can move at a 
constant velocity \cite{1D}. Bright solitons have been studied and observed in 
different quantum  and classical systems, such as,  
in nonlinear optics \cite{nlo} and Bose-Einstein condensates (BEC) 
\cite{expt}, and in water waves. Usually,  1D bright solitons are  
analytic with energy and momentum conservation which guarantees  mutual elastic 
collision  with shape preservation.  Following a theoretical 
suggestion \cite{theo},
quasi-1D solitons have been realized \cite{expt} in a  cigar-shaped BEC 
with strong  transverse confinement.  
Due to a collapse instability  such a soliton in a stationary state cannot be realized \cite{1D,nlo}
in  three-dimensions (3D) for attractive  
interaction. However, a dynamically stabilized non-stationary  3D state can be achieved \cite{sadhan}.

The BEC bright solitons 
 are usually  studied  theoretically  with the    
mean-field Gross-Pitaevskii (GP) equation \cite{gp}. 
It was   shown  that the  inclusion of a perturbative
Lee-Huang-Yang (LHY) interaction \cite{Lee, axel,Petrov1} or of a repulsive  three-body interaction
\cite{adhikari}, both valid  in the mean-field domain,  generating   higher-order repulsive 
nonlinear terms   in the  GP equation   compared to the cubic nonlinear two-body  
terms, can avoid the collapse and thus form a self-bound  3D BEC state. 
Petrov \cite{Petrov1} showed that a  self-bound   binary BEC state can be formed in 3D 
in the presence of  intra-species repulsion with the  LHY  interaction and an inter-species attraction. 
Under the same setting a self-bound binary boson-fermion state can be formed in 3D  \cite{arxiv}. 
   A self-bound state can also be realized in a multi-component spinor 
BEC with spin-orbit or Rabi interaction \cite{malomed}.
Self-bound states were observed in        dipolar $^{164}$Dy 
\cite{Kadau}  and 
$^{166}$Er \cite{Chomaz}
BECs.   The formation of self-bound dipolar  states
was  explained by means of the   LHY  interaction \cite{Wachtler1,Schmitt,Barbut,Bisset}.
More recently, a  self-bound binary BEC  
of two hyper-fine states of $^{39}$K in the presence of     inter-species attraction  and  
intra-species repulsion  has been observed \cite{Cabrera,Cheiney,inguscio} and theoretically studied by including the LHY interaction.

The  LHY interaction \cite{Wachtler1,Schmitt,Barbut,Bisset,Cabrera,Cheiney,inguscio,ing2,fish}, used   in stabilizing a 3D self-bound state, is perturbative in nature valid for small values of the gas parameter $x\equiv an^{1/3}$ in the mean-field domain, where $a$ is the scattering length and $n$ the density, and hence has limited validity.
For a complete  description of  the problem,  a realistic non-perturbative  beyond-mean-field (BMF)  interaction should be employed. We use a realistic analytic non-perturbative BMF interaction valid for both small ($x\ll 1,$ weak coupling) and large ($ x\gg 1,$ strong coupling) values of the gas parameter which reproduces the result of a microscopic multi-channel calculation of the BMF interaction.  For weak coupling, this non-perturbative BMF interaction reduces to  the LHY interaction and for strong coupling it has the proper unitarity limit.  The LHY interaction leads 
to a higher-order repulsive quartic non-linearity in the dynamical model compared to the 
cubic non-linearity of the GP equation.% arising from the two-body contact interaction in the Hartree approximation.  
 Without this higher-order 
term, an attractive BEC has an infinite negative  energy at the center leading to 
a collapse of the system to the center. The higher-order repulsive LHY  interaction leads to an infinite positive energy at the center and stops the collapse.
 The realistic non-perturbative BMF interaction does not have such a term except in the extreme weak-coupling limit $(x\lessapprox  0.01)$.
  For most values of coupling ($+\infty >x>0.1$), such a higher-order nonlinear term is absent in the non-perturbative  realistic  two-body
 BMF interaction and a self-bound state cannot be formed. Similar deficiency of the perturbative  LHY model in describing self-bound BEC states has been recently 
pointed out  for a binary BEC mixture \cite{rec} and for a dipolar BEC \cite{pfau}.
 In these cases a three-body   interaction can possibly stabilize a self-bound BEC state \cite{adhikari,arxiv}.
We demonstrate our point of view in a study of self-binding in   a binary  $^{39}$K  BEC in two different hyper-fine states.  We derive  the nonlinear model
equation  with the non-perturbative BMF interaction   and solve it 
numerically. In addition, we consider a variational approximation {\cite{vari}  to 
this  model in the weak-coupling limit with the perturbative  LHY interaction for a qualitative understanding.

\section{Analytical Formulation}

 The two-body   interaction  energy density ${\cal E}$ (energy per unit volume) of a homogeneous dilute weakly repulsive Bose gas including the  LHY   interaction \cite{Lee}   is given by {\cite{Petrov1,adhikari}} 
\begin{eqnarray}
{\cal E}(n,a) &=&  \frac{U  n^2}{2}
\left( 1+\frac{2\alpha\sqrt{na^3}}{5}\right), \quad \alpha=\frac{64}{3\sqrt \pi}  ,\label{Energy}\\
&=&  \frac{U  n^2}{2}  + \frac{8m^4}{15 \pi^2 \hbar^3} c^5,
\end{eqnarray}
where $U=4\pi \hbar^2a/m$ is the strength of two-body interaction,   
 $m$ is the mass of an atom, $c=\sqrt{Un/m} $  is the speed of  sound in a single-component BEC \cite{gp,Pethick}.     A localized BEC of $N$ atoms  with  number density $n= N|\psi({\bf r},t)|^2$, where $ \psi({\bf r},t)$ is the wave function at time $t$ and space point $\bf r$ and  with interaction (\ref{Energy}) is described by   
the following time-dependent mean-field  NLS equation \cite{Petrov1,Cabrera,AS} with the perturbative LHY correction:  
\begin{align} \label{nls}
{\mathrm i}\hbar\frac{\partial \psi({\bf r},t)}{\partial t}  &= \left[ -\frac{\hbar^2 \nabla^2}{2m} + \mu(n,a)\right] \psi({\bf r},t), \\
\mu(n,a) &= \frac{\partial {\cal E}}{\partial n}= \frac{4\pi \hbar^2 
an}{m}\left(1 +\frac{\alpha}{2} { \sqrt{na^{3}} }\right),\label{MU}
\end{align}
with normalization $\int |\psi|^2 d{\bf r}=1$
where $\mu(n,a)$ is the chemical potential of the homogeneous Bose gas.  

A convenient dimensionless form of Eqs. (\ref{Energy}), (\ref{nls})  and (\ref{MU}) can be obtained with the scaled variables ${\bf r}'={\bf r}/l_0$, $a'=a/l_0$, $n'= n l_0^3,$ $\psi' = l_0^{3/2} \psi, $   $t'=\hbar t/ml_0^2$,  $\mu'=\mu ml_0^2/\hbar^2 $, etc., with $l_0$ a length scale :
 \begin{align}
\label{energy1}
{\cal E}(n,a) &= 2\pi a n^2   \left(1 +\frac{2\alpha}{5} { \sqrt{na^{3}} }\right),
\\
 \label{NLS}
{\mathrm i}\frac{\partial \psi({\bf r},t)}{\partial t}  &= \left[ -\frac{ \nabla^2}{2} + \mu(n,a)\right] \psi({\bf r},t), \\
\mu(n,a) &= {4\pi  
an}\left(1 +\frac{\alpha}{2} { \sqrt{na^{3}} }\right),\label{mu}
\end{align}
where we have dropped the prime from the transformed variables and $n=N|\psi|^2, \int d{\bf r}|\psi|^2=1$.
%The mass $m$ and reduced Planck constant $\hbar$ only affects the transformed time and energies and does not affect the transformed length and density which are detemined by the length scale $l_0$. This is possible in the absence of trap.  

The $\alpha$-dependent terms in Eqs. (\ref{Energy}) and (\ref{MU}) are the {\it lowest-order}
  {\it perturbative}
LHY corrections to the mean-field   energy density and chemical potential and 
like all perturbative corrections  have    limited validity  for small values of the gas parameter 
$x\equiv (n^{1/3}a) \ll 1$.  For larger values of $x$ the higher-order corrections become important, and  specially  as $a\to \infty$ at unitarity, the $\alpha$-dependent terms in Eqs. (\ref{Energy}) and (\ref{MU}) diverge even faster than the mean-field  term proportional to scattering length, while the energy density $\cal E$ and the chemical potential $\mu$ remain finite at unitarity.  At unitarity $a\to \infty$,  the bulk chemical potential $\mu(n,a)$ cannot be a function of the scattering length $a$ and by dimensional argument  should instead be proportional to $n^{2/3}$, e. g., \cite{AS}
\begin{equation}\label{unimu}
\lim_{a\to \infty} \mu(n,a)  = \eta n^{2/3},
\end{equation}
where $\eta$ is a universal constant.  The corresponding expression for energy density at unitarity is \cite{AS}
\begin{equation}\label{unien}
\lim_{a\to \infty} {\cal E}(n,a)  = \frac{3}{5}\eta n^{5/3},
\end{equation}
with the property $\mu(n,a)= \partial {\cal E}(n,a)/ \partial n$.

 The following analytic BMF  {\it non-perturbative} chemical potential  with a single parameter $\eta$  valid for both small and large values of gas parameter $x$  is   useful for phenomenological application, for example, in the formation of a self-bound state in a binary BEC  \cite{AS}:
\begin{eqnarray} \label{ASmu}
\mu(n,a) &=& n^{2/3}f(x) \equiv  n^{2/3} \frac{4\pi(x+\alpha x^{5/2})}{1+\frac{\alpha}{2}x^{3/2}+ \frac{4\pi\alpha}{\eta} x^{5/2}},   
\end{eqnarray}
where $f(x)$ is a universal function with properties $\lim_{x\to 0} f(x)= 4 \pi x +2\pi \alpha x^{5/2}$ and $\lim_{x\to \infty} f(x)= \eta.$ Consequently,
for small and large values of $x$ the chemical potential (\ref{ASmu}) reduces to the limits
(\ref{mu})  and (\ref{unimu}), respectively.  We will call this model  the crossover model,  as it is valid for all coupling along the crossover from weak coupling to extreme strong coupling (unitarity). 
Similarly, an analytic expression for energy density with the correct weak- and strong-coupling limits 
(\ref{energy1})  and (\ref{unien})   is
\begin{eqnarray} \label{ASen}
{\cal E}(n,a) &=& n^{5/3} \frac{2 \pi(x+\frac{4 \alpha}{5} x^{5/2})}{1+\frac{2\alpha}{5}x^{3/2}+ \frac{8\pi\alpha}{3\eta} x^{5/2}}.   
\end{eqnarray}

\begin{figure}[t]

\begin{center}

\includegraphics[trim = 0.225cm -0.1cm 0cm 0cm, clip,width=.49\linewidth,clip]{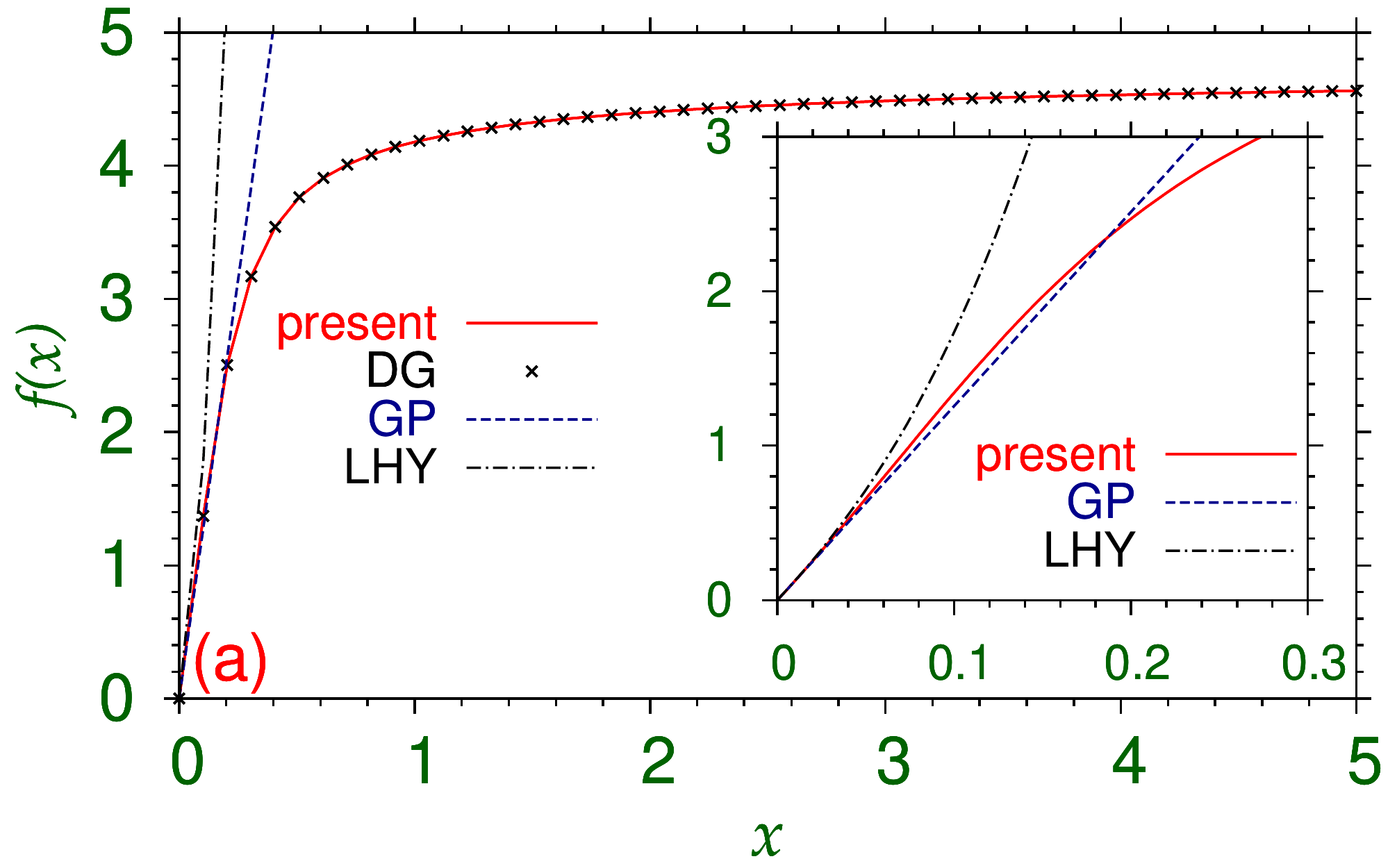} 
\includegraphics[trim = 0.225cm -0.1cm 0cm 0cm, clip,width=.49\linewidth,clip]{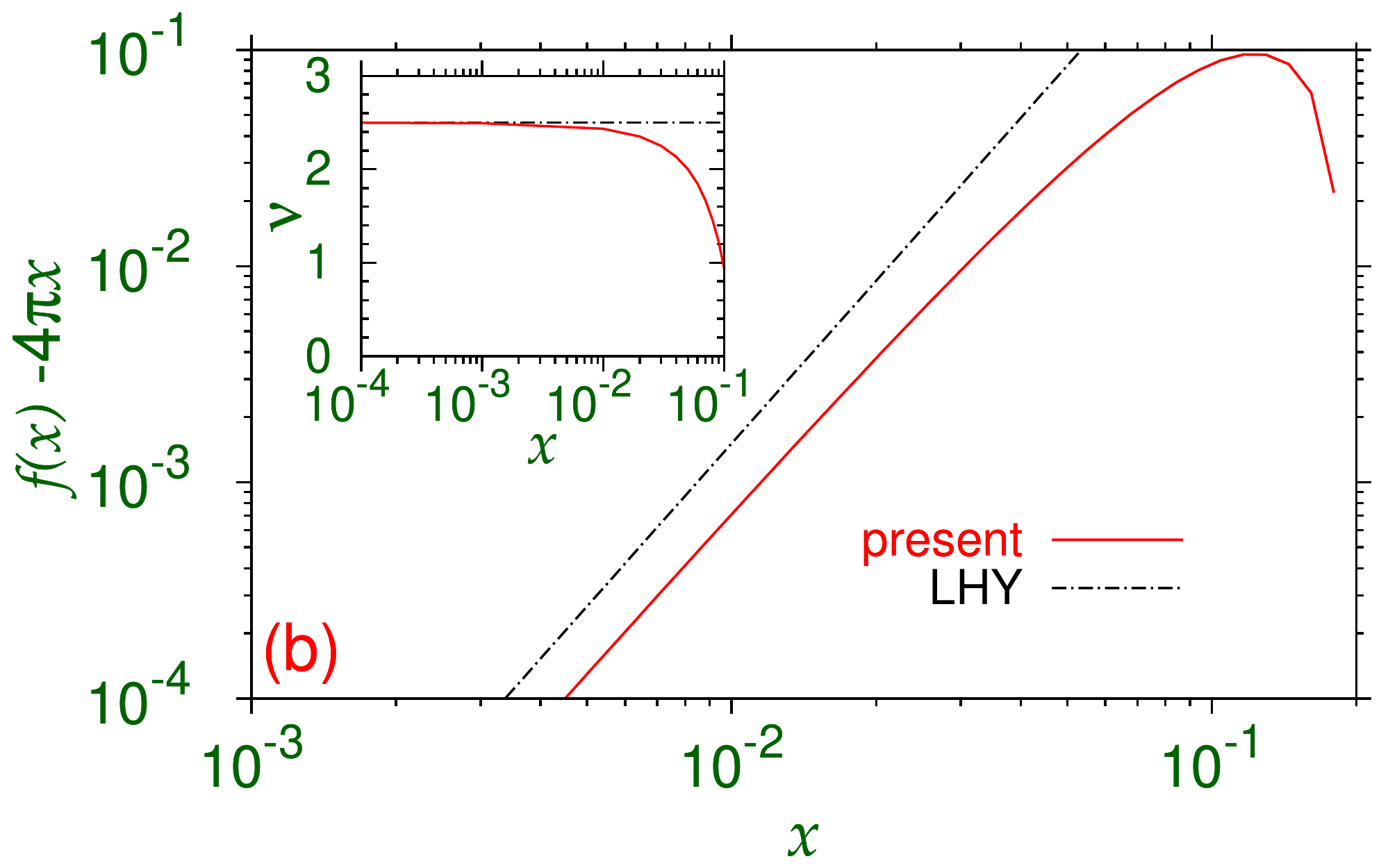} 
\caption{(a)  The crossover analytic function $f(x)$ of Eq. (\ref{ASmu}) with $\eta=4.7$
compared with the realistic multi-channel Hartree calculation of DG \cite{DG}, the LHY approximation 
(\ref{mu}), and the GP limit $f(x)=4\pi x$.  The inset shows the same for small $x$. 
(b)  A log-log plot of the function $[f(x)-4\pi x]$ versus $x$ for the crossover  BMF interaction (\ref{ASmu}) and its LHY approximation 
(\ref{mu}). The slope of this plot determines the exponent $\nu$ of the scaling relation $[f(x)-4\pi x]\sim x^\nu$. The inset displays the exponent $\nu$ versus $x$ for the two functions. 
 The plotted quantities in this and following figures are dimensionless.}  
\label{fig1}
\end{center}
\end{figure}

Although there are no precise experimental estimates   of the universal constant $\eta$ and the universal function $f(x)$,
 there are several microscopic calculations of the same \cite{DG,micro}. The most recent multi-orbital microscopic Hartree calculation of $\mu(n,a)$ by Ding and Greene (DG) along the weak- ($x\ll 1$) to strong-coupling ($x\gg 1$) crossover yields $\eta = 4.7$ \cite{AS,DG}. Other  microscopic calculations  \cite{micro} yield $\eta$ in the range from 3 to 9.
In Fig. \ref{fig1}(a)  we compare the crossover function  $f(x)$ of Eq. (\ref{ASmu}) for $\eta=4.7$ with the same obtained by DG, as well as with the  LHY approximation $f(x)=4\pi x+ 2\pi \alpha x^{5/2}$ and the mean-field GP value $f(x)=4\pi x$.   We see that the crossover function $f(x)$ is in full agreement with the microscopic calculation of DG for all $x$. 
The mean-field GP and the  LHY approximations 
diverge for large $x$ and these two perturbative results are valid in the weak-coupling limit 
($x\ll 1$). This is further illustrated in the inset of Fig. \ref{fig1}(a) where  we exhibit these functions for 
small $x$. Even for small $x$ ($x\ll 1$),   $f(x)$  of the LHY model could be very different from the  non-perturbative interaction (\ref{ASmu}) and the GP result. Nevertheless, the higher-order $x^{5/2}$ non-linearity in the LHY 
model will always stop the collapse and allow the formation of a self-bound state, whereas the crossover interaction   (\ref{ASmu})
will  stop the collapse only for very small values of $x$, where it   tends to the LHY model.   To see the higher-order non-linearity in the function $f(x)$ of the crossover  model  (\ref{ASmu}), responsible for arresting the collapse,    more clearly, we illustrate in Fig. \ref{fig1}(b)  the correction to the GP model 
$[f(x)-4\pi x]$ calculated using the crossover  and the LHY interactions versus $x$ in a log-log plot. The slope of this plot gives 
the power-law exponent $\nu$ of the scaling relation $[f(x)-4\pi x] \sim x^\nu$. A large $\nu$ ($> 1$) is required to stop the collapse. 
 In the inset of this figure we plot the exponent $\nu$ versus $x$.   We find that for the LHY model $\nu =2.5$ for all $x$, but for the  realistic crossover  model  (\ref{ASmu}) $\nu \approx 2.5$ for $x\lessapprox 0.01$ and then rapidly 
reduces and has the value $\nu \approx 1$ for $x \approx 0.1$ still for weak-coupling ($x<1$). Hence the LHY model is realistic only for $x\lessapprox 0.01$.  In the study of the formation of a self-trapped binary BEC the LHY model should be applied within the domain of its validity.

For a binary BEC, there are two speeds of sound $c_i$ \cite{Pethick} for the two components denoted by $i=1,2$ of identical mass $m$, and the energy density can be written as \cite{xyz}
\begin{align}
{\cal E}(n_i,a_i) = \sum_i  \left[  \frac{U_in_i^2}{2}  + \frac{8m^4}{15\pi^2 \hbar^3}  c_i^5   \right] +U_{12}n_1 n_2
  \end{align}
where $n=n_1+n_2$. Here the two-body
intra- and inter-species interaction strengths are $ U_i = 4\pi \hbar^2 a_i/m$ with $a_i$ the  
intra-species scattering length for species $i$, and $U_{12} = 4\pi \hbar^2 a_{12}/m$, where $a_{12}$ is the
  inter-species scattering length, respectively. The two speeds of sound for this system are \cite{Pethick,Gladush}
\begin{align}
c_i\equiv c_{\pm} = \sqrt {\frac{\sum_i U_in_i \pm \sqrt{(U_1n_1-U_2n_2)^2+4n_1n_2 U_{12}^2}}  {2m}}.
  \end{align}
If $a_{12}\approx -\sqrt{a_1a_2}$  for repulsive intra-species and attractive inter-species interactions,
$c_-\approx 0$ and $c_+= \sqrt{(U_1n_1+U_2n_2)/m}.$ Then the mean-field energy with LHY interaction for the binary system becomes \cite{sandeep}
\begin{align}
{\cal E}(n_i,a_i)  = \sum_i    \frac{U_in_i^2}{2}  +U_{12}n_1 n_2  +  \frac{8m^4}{15\pi^2 \hbar^3} 
\left( \frac{\sum_i U_in_i}{m}  \right) ^{5/2}
  \end{align}

We will see that even for weak-coupling ($x<1$), the results for the self-bound binary BEC mixture obtained using the LHY interaction may not be realistic compared to the  non-perturbative  crossover  BMF
interaction (\ref{ASmu}). To demonstrate this we consider a symmetric binary BEC mixture of components $i=1,2$ with an equal number of atoms $N_1=N_2=N/2$ in the two components, where $N_i $ is the number of atoms in component $i$ and   $N$ total number of atoms. The binary mean-field  equations with the  LHY interaction in dimensionless units ($\hbar = m=1$) are \cite{Petrov1,sandeep}
 \begin{eqnarray}
{\mathrm i}\frac{\partial \psi_1}{\partial t}  = \biggr[-\frac{ \nabla^2}{2} + 2\pi N a_1|\psi_1|^2 + 2\pi Na_{12}|\psi_2|^2  +\frac{\alpha}{\sqrt 2} \pi N^{3/2}a_1   
 \Big(\sum_i a_i |\psi_i|^2 \Big)^{3/2}\biggr]\psi_1,\label{GPE1}\\
{\mathrm i}\frac{\partial \psi_2}{\partial t}  = \biggr[-\frac{ \nabla^2}{2} +   2\pi Na_2|\psi_2|^2 + 
2\pi N a_{12}|\psi_1|^2 +\frac{\alpha}{\sqrt 2} \pi N^{3/2}a_2  
 \Big(\sum_i a_i |\psi_i|^2 \Big)^{3/2}
\biggr]\psi_2,\label{GPE2}
\end{eqnarray}
where $\psi_1$ and $\psi_2$ are wave functions of the two components,   normalized as $\int |\psi_i({\bf r},t)|^2 d{\bf r}=1$. The nonlinear terms in Eqs. (\ref{GPE1})  and (\ref{GPE2}) are the chemical potentials $\mu_i= 
\partial {\cal E}(n_i,a_i) /\partial n_i$.
 
 If we further take $a_1=a_2=a$, $a_{12}=-a-\delta$  and $\psi_1=\psi_2=\psi$, then Eqs. (\ref{GPE1}) and (\ref{GPE2}) become \cite{sandeep}
  \begin{align} \label{eff}
{\mathrm i}\frac{\partial \psi({\bf r},t)}{\partial t}  = &\biggr[-\frac{ \nabla^2}{2} 
- 4\pi N{\Big(a+\frac{\delta}{2}\Big)}|\psi|^2  +  4\pi N a|\psi|^2\biggr(1+\frac{\alpha}{2}
 \sqrt{Na^3}|\psi|\biggr) \biggr]\psi({\bf r},t),
\end{align}
 where $\psi({\bf r},t)$ is the wave function of the self-bound state containing $N$ atoms and normalized as $\int |\psi({\bf r},t)|^2 d{\bf r}=1$. Equation (\ref{eff}) is a mean-field  equation, with the LHY interaction,  satisfied by the self-bound state and is identical to  Eq. (1) of Refs. \cite{Cabrera,Cheiney} and equivalent to Eq. (10) of Ref. \cite{Petrov1} for small $\delta$ .  This equation is
 the same as  the single-component mean-field equation (\ref{NLS}) 
with the LHY approximation
with an additional attractive nonlinear term $- 4\pi N{(a+\delta/2)}|\psi|^2 $. In both Eqs. (\ref{NLS})
and (\ref{eff})
 we have $N$ atoms with scattering length 
$a$. The extra 
attractive non-linearity in Eq. (\ref{eff})  is of the GP type with scattering length $-(a+\delta/2)$.   
A self-bound binary state is possible for a positive 
value of $\delta.$

This close resemblance of the  perturbative  BMF  equation (\ref{eff})  with the NLS equation 
(\ref{NLS}) allows us to include the realistic BMF interaction (\ref{ASmu}) to the chemical potential. The  so-corrected non-perturbative BMF crossover model  for the self-bound state  is 
  \begin{align} \label{present}
{\mathrm i}\frac{\partial \psi({\bf r},t)}{\partial t}  = &\biggr[-\frac{ \nabla^2}{2} 
- 4\pi N{\Big(a+\frac{\delta}{2}\Big)}|\psi|^2  +\mu(n,a) \biggr]\psi({\bf r},t),
\end{align}
where $\mu(n,a)$ is  now given by Eq. (\ref{ASmu}) with  $n=N|\psi|^2$ and $\int d{\bf r}|\psi({\bf r},t)|^2 =1$. 

In the following we make a comparative study of  the formation of a self-bound binary state using the LHY and crossover  models  (\ref{eff}) and (\ref{present}). The result obtained from these two models can be widely different even in the weak-coupling limit  $x< 1.$  A positive $\delta$ usually leads to a self-bound state in the perturbative LHY model  (\ref{eff}), whereas the realistic non-perturbative model (\ref{present}) leads to a  self-bound state  only for very small values of the gas parameter $x$ ($x\lessapprox 0.01$), where  the exponent $\nu$ is close to 2.5, viz., Fig. \ref{fig1}: for medium values of $x$ in the weak-coupling limit a self-bound state may not be realized 
in the crossover model  (\ref{present}). The reason is that for an added net attraction the collapse in the LHY model (\ref{eff}) will be stopped by the higher-order quartic LHY non-linearity. In the realistic BMF model (\ref{present}) this higher-order non-linearity exists only for extreme weak coupling, viz. Fig. \ref{fig1}(b).  How a higher-order non-linearity stops the collapse is illustrated next making  a variational approximation to the  LHY model (\ref{eff}).

The Lagrangian density for Eq. (\ref{eff}) is \cite{sandeep}
  \begin{align}
{\cal L} =      \frac{N}{2} \Big\{  {\mathrm i}  \big(  \phi   \phi^{*'} - \phi^*
 \phi ^{'}\big)+  \big| \nabla \phi\big|^2 
 \Big\}
-\pi N^2 \delta  |\phi|^4 
+ \frac{256 \sqrt \pi}{15} \big( N a\big)^{5/2} |\phi|^5,
 \label{Lagrangian2}
\end{align}
where the prime denotes time derivative.
This Lagrangian density with the time-derivative terms   set equal to zero is the same as the stationary energy density given by Eq. (1) of Ref. \cite{Cabrera} under the condition  $a_1=a_2\equiv a$: 
 \begin{align}
{\cal E}=      \frac{N}{2}     \big| \nabla \phi\big|^2 
-\pi N^2 \delta  |\phi|^4  
+ \frac{256 \sqrt \pi}{15} \big( N a\big)^{5/2} |\phi|^5.
 \label{energy}
\end{align}

A Lagrange variational approximation to Eq.  (\ref{eff}) can be performed with the following 
 variational {\em ansatz} {\cite{vari}} 
\begin{eqnarray}
\phi ={\pi ^{-3/4} w^{-3/2}} {\exp \Big({-\frac{r^2}{2 w^2}+\text{i} \kappa r^2}\Big)},\label{gaussian_ansatz}
\end{eqnarray}
where $w$ is the width and $\kappa$ the chirp. Using this {\em ansatz}, the Lagrangian density   (\ref{Lagrangian2}) can be integrated over all space to yield the   Lagrangian functional  \cite{sandeep}
\begin{align}\label{lagr}
L \equiv  \int {\cal L}d{\bf r} ={3}\kappa ^2 w^2N+\frac{3}{2} w^2 \kappa 'N+\frac{3 N}{4 w^2}
- \frac{\pi N N \delta }{2 \sqrt{2} \pi ^{3/2} w^3} 
+ \frac{512 \sqrt{{2}} \left( Na\right){}^{5/2}}{75\sqrt 5  \pi ^{7/4} w^{9/2}}  .
\end{align}
The total energy of a  self-bound stationary binary BEC is  
 \begin{align}\label{ener}
E\equiv  \int {\cal E}d{\bf r}  =\frac{3 N}{4 w^2}
- \frac{\pi N N \delta }{2 \sqrt{2} \pi ^{3/2} w^3} 
+ \frac{512 \sqrt{{2}} \left( Na\right){}^{5/2}}{75\sqrt 5  \pi ^{7/4} w^{9/2}}  .
\end{align}
In the absence of the last term of Eq. (\ref{ener}) with the LHY contribution, 
the  energy $E$  of a stationary state  tends to $-\infty$ as $w\to 0$ signaling a collapse instability. However, in the presence of the LHY
interaction,  the energy at the center ($w=0$) becomes infinitely large $(+\infty)$ and hence a collapse is avoided. 
The Euler-Lagrange equations of the Lagrangian (\ref{lagr}) for variables $\gamma  \equiv \kappa, w$, 
\begin{equation}
\frac{\partial }{\partial t}\frac{\partial L}{\partial \gamma '}-\frac{\partial L}{\partial \gamma }=0,
\end{equation}
lead to  
\begin{align}
 w'' =\frac{ 1}{ w^3}- \frac{  N \delta }{\sqrt{2\pi }  w^4 }  +
\frac{512    \sqrt{2N^3 a^5}}{25 \sqrt 5 \pi ^{7/4} w^{11/2}} , \label{EL_eq}
\end{align}
which describes the variation of the width of the self-bound state with time.  The width of a stationary state is obtained by setting $w''=0 $ in Eq. (\ref{EL_eq}).

\begin{figure}[t]
\begin{center}
\includegraphics[trim = 0cm 0cm 0cm 0cm, clip,width=.6\linewidth,clip]{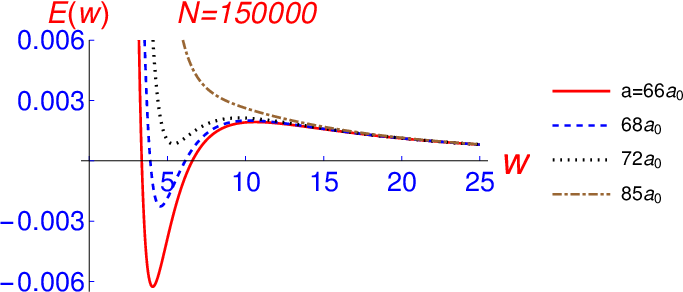}  

\caption{  Variational energy per atom $E(w)=E/N$, viz. Eq. (\ref{ener}), as a function of width for 
$N=150000,$  $a=66a_0, 68a_0, 72a_0,$  and $85a_0$, and $\delta =4a_0$.  The negative-energy minima for  $a=66a_0, 68a_0$ correspond to a stable state and the positive-energy minimum for $a=72a_0$ denotes a meta-stable state. The minimum has disappeared for $a=85a_0$ indicating an unbound configuration.  }
\label{fig2}
\end{center}
\end{figure}

\begin{figure}[t]
\begin{center}
\includegraphics[trim = 0cm 0cm 0cm 0cm, clip,width=.47\linewidth,clip]{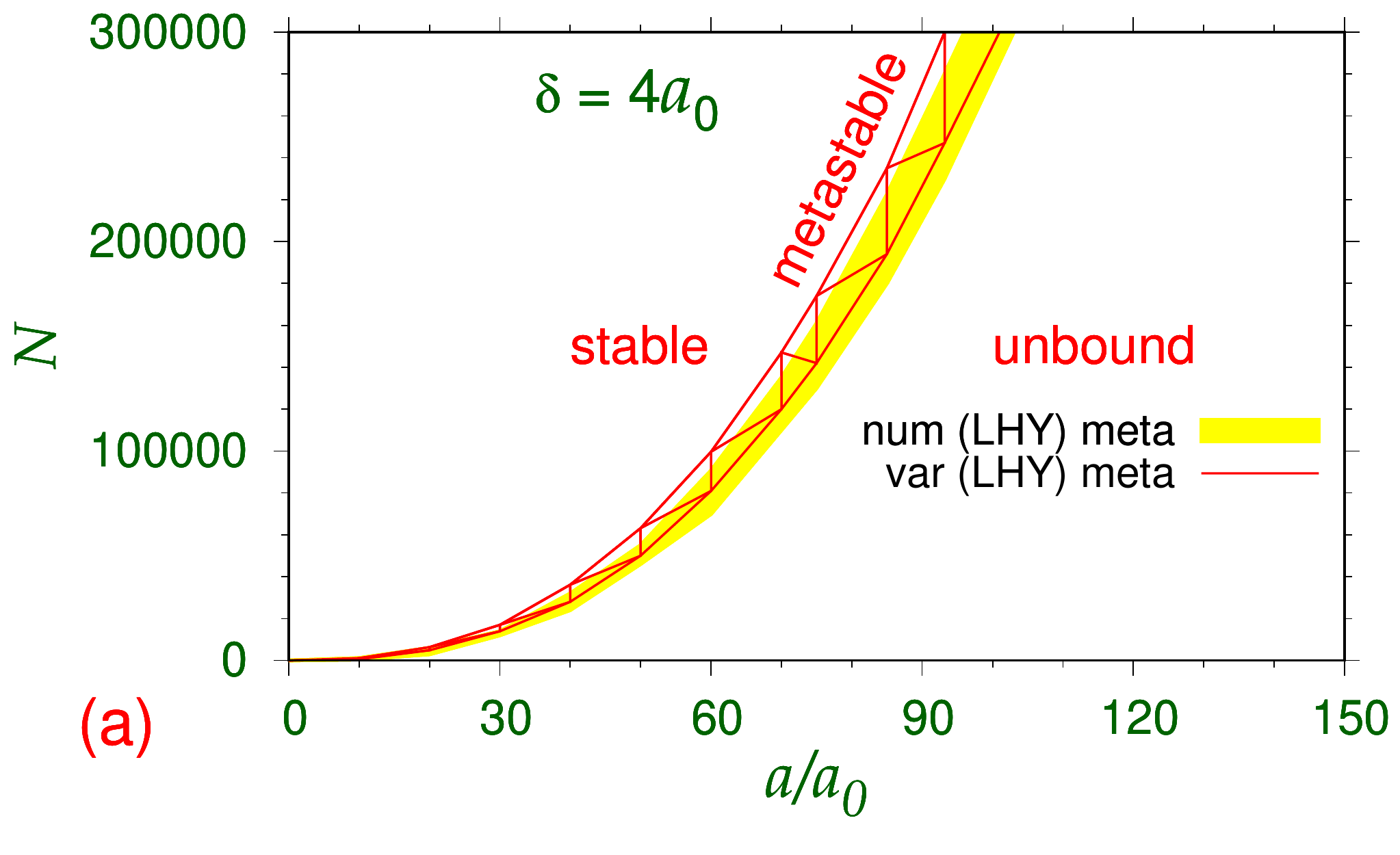} 
\includegraphics[trim = 0cm 0cm 0cm 0cm, clip,width=.47\linewidth,clip]{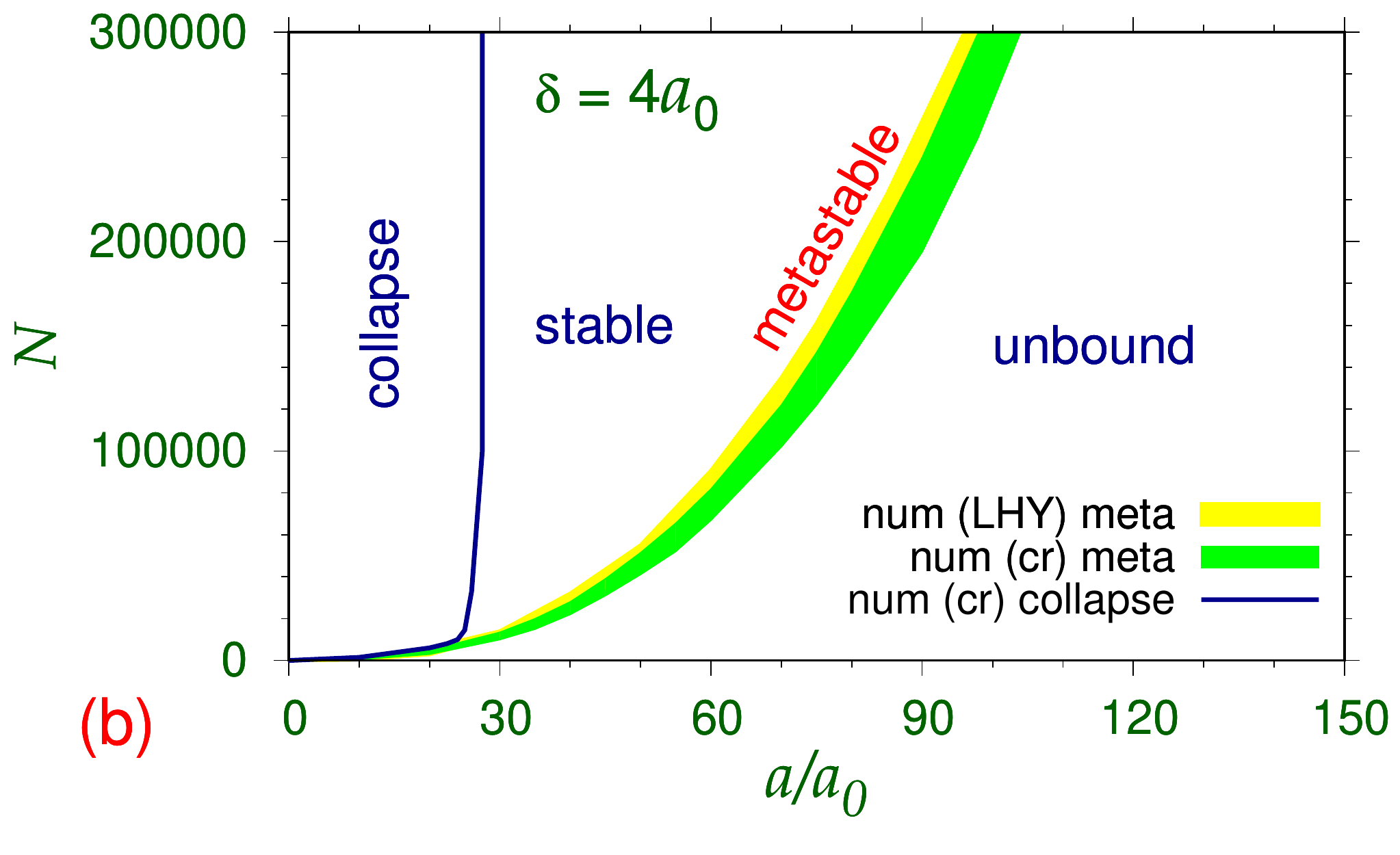} 
\includegraphics[trim = 0cm 0cm 0cm 0cm, clip,width=.49\linewidth,clip]{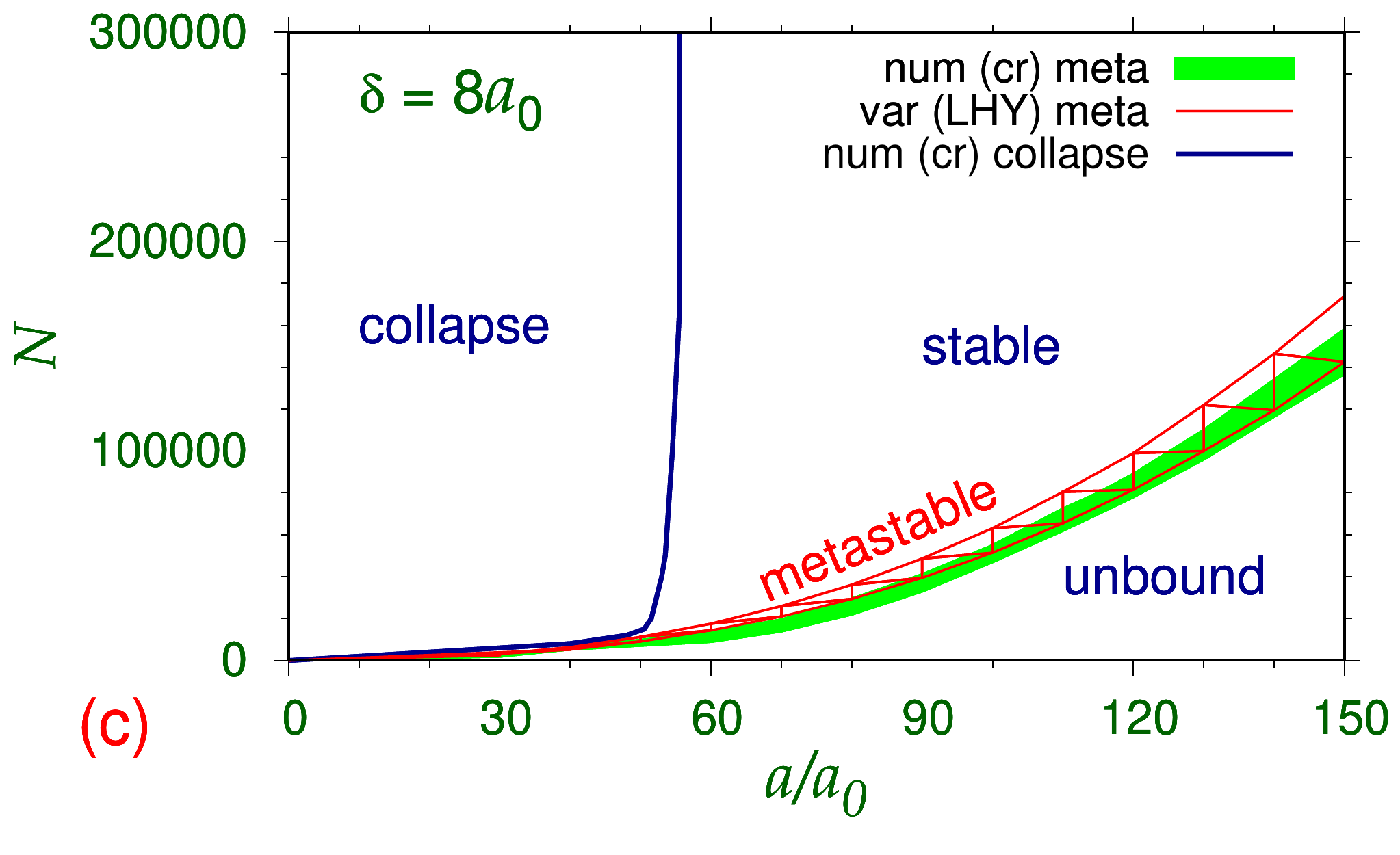}
\caption{  Stable, meta-stable (meta), unbound and collapse regions highlighted in the $N-a$ phase plots. Results of
(a) numerical (num) calculation and  variational (var) approximation based on the LHY model (\ref{eff}) with $\delta=4a_0$, (b) numerical  calculation based on the  LHY and crossover (cr) models (\ref{eff}) and (\ref{present})  with  $\delta=4a_0$,
(c)   numerical calculation and  variational approximation based on the   LHY and crossover  models (\ref{eff}) and (\ref{present}) with $\delta=8a_0$. The region marked ``collapse" denotes collapse of the system only for the non-perturbative BMF crossover model (\ref{present}).}
\label{fig3}
\end{center}
\end{figure}

\section{ Numerical Result}
 The effective nonlinear equation (\ref{present}) does not have analytic solution and different numerical methods, for example, split-step Crank-Nicolson \cite{am} or pseudo-spectral \cite{psp} method, are employed for its solution.   
We use split time-step Crank-Nicolson method to solve Eq,  (\ref{present})  numerically  
  \cite{Muruganandam}.  
The minimum-energy  ground-state solution for the self-bound state
is obtained by evolving the trial wave functions, chosen to be Gaussian, in imaginary
time $\tau = i t$   as is proposed in Ref. \cite{am}. 
The numerical results  of the models for stationary self-bound state  (\ref{eff}) and (\ref{present}) in a spherically symmetric   configuration  are 
presented and critically contrasted  in the following  in spherical coordinate $r$.  
  We consider  spatial and time steps,  to solve the NLS equations in imaginary-time propagation, as small as   
$r=0.0125$ and $t=10^{-5}$ and take the length scale $l_0=1$ $\mu$m throughout this study.  A small space step is needed to find out the possible collapse of the 
self-bound state when its size  becomes very small. 

 We consider a binary self-bound state consisting of   $| m_F= -1\rangle$  and $| m_F= 0\rangle$ hyper-fine states of 
$^{39}$K with an equal number $N/2$  of atoms in the two states, which has been realized experimentally \cite{Cabrera,Cheiney}.  The intra-species  scattering lengths of  
the two components can be varied by a Feshbach resonance \cite{fesh} resulting in a variation of   
 the scattering length $a=\sqrt{a_1a_2}$..   
These scattering lengths are kept  quite close to each other and we take $a= \sqrt{a_1a_2}\approx a_1 \approx a_2$.   The inter-species scattering length $a_{12}$    is taken as $a_{12}= -a -\delta$ with 
$\delta =4a_0$ and $8a_0$ covering the range of $\delta$ values  considered  in the experiment \cite{Cabrera}: $\delta=  2.4a_0, 3.2a_0, 3.8a_0, 4.4a_0, 5a_0$ and $5.5a_0$  with $a_0$ the Bohr radius.   In this {\it model study} we will vary $a$ and contrast the results obtained with models (\ref{eff}) and (\ref{present}).

The variational result for energy with the LHY interaction (\ref{ener})  confirm the existence of energetically meta-stable as well as stable  self-bound states. To illustrate the
distinction between a meta-stable and a stable  state, the variational energy per atom  $E(w)\equiv E/N$ of  binary self-bound states are displayed in Fig. \ref{fig2}
 as a function of the variational width $w$   for $N=150000$,   $\delta =4a_0,$ and $ a=66a_0, 68a_0, 72a_0,$  and $85a_0$.
In Fig. \ref{fig2}, a meta-stable  state corresponds to a curve with a local minimum in the energy (at a positive energy), viz. $a=72 a_0$,
whereas   a stable state corresponds to the curve with a global minimum (at a negative energy), viz. $a=66 a_0, 68a_0$. The energy, as $w\to \infty$, is zero.   An unbound  state corresponds to a curve with no minimum, viz. $a=85a_0$ in Fig. \ref{fig2}. All states for $ a<66a_0 $ are stable, and those with $a>85a_0 $ are unbound.

\begin{figure}[t]
\begin{center}
\includegraphics[trim = 0cm 0cm 0cm 0cm, clip,width=.49\linewidth,clip]{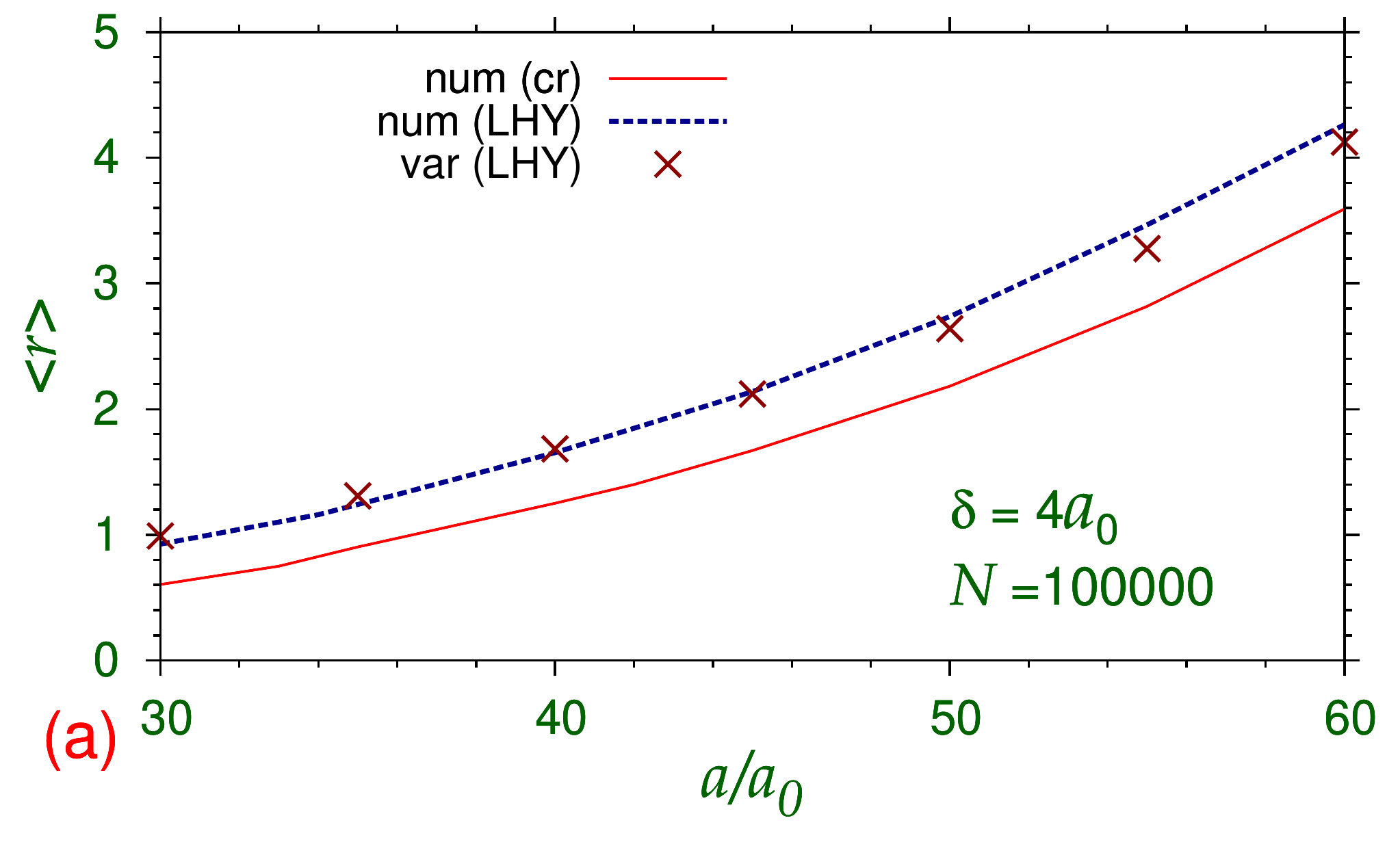} 
\includegraphics[trim = 0cm 0cm 0cm 0cm, clip,width=.49\linewidth,clip]{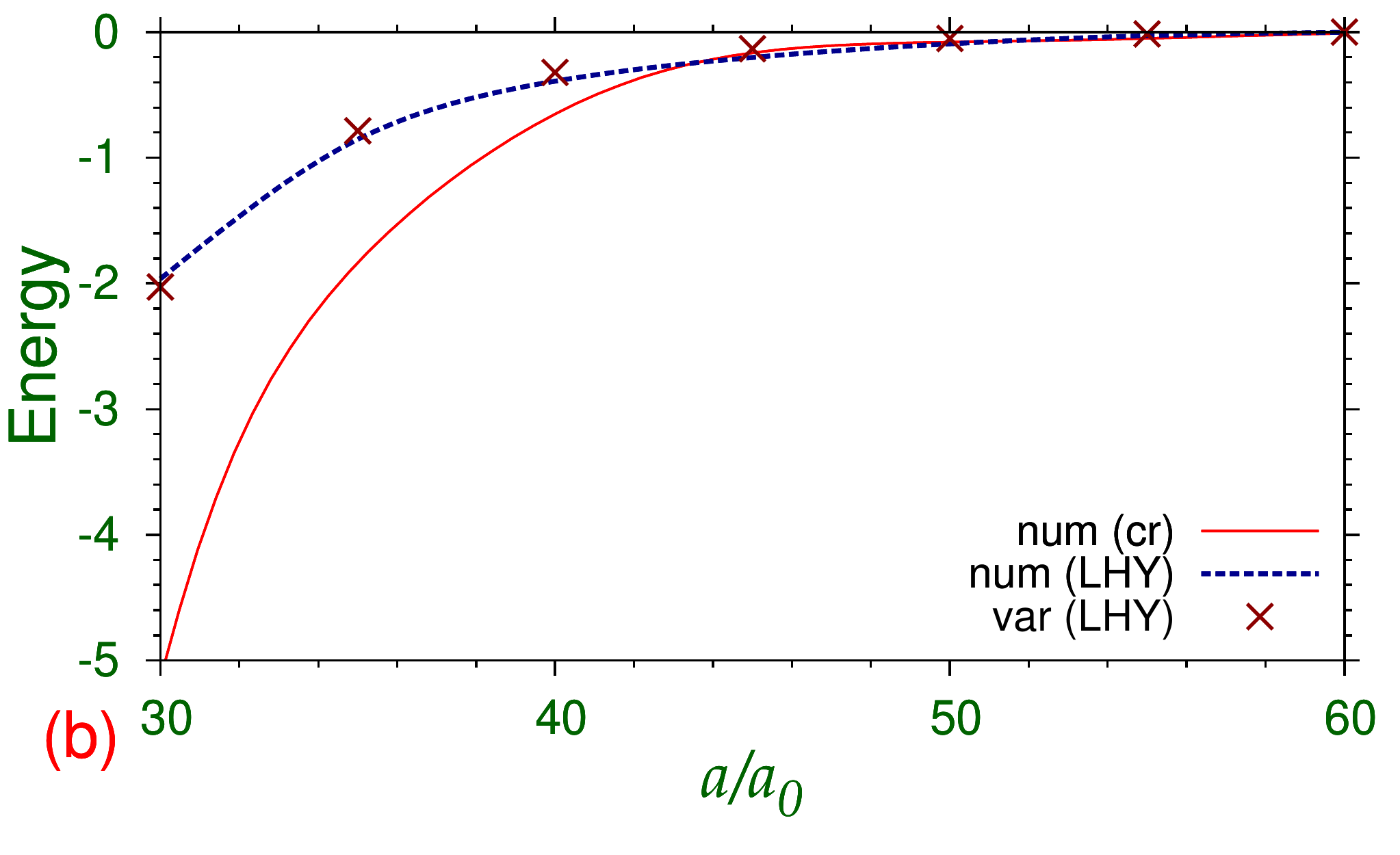}  
\caption{    Results of
 numerical (num) calculation and  variational (var) approximation to (a) root mean square radius and (b) energy
$E/N$ 
based on the  LHY model (\ref{eff}) and crossover (cr) model (\ref{present}) with $\delta=4a_0$ and $N=100000$.}
\label{fig4}
\end{center}
\end{figure}

 The variational phase plot in the $N-a$ plane 
 of the formation of a self-bound binary BEC state  with the perturbative   LHY model (\ref{eff}), while
keeping  $\delta$  fixed at $4a_0$, is illustrated  in Fig. \ref{fig3}(a).   The variational phase plot is obtained by exploring the minimum of variational energy (\ref{ener}).   The region marked stable corresponds to global minima of energy whereas meta-stable to local minima of energy. In the region marked  unbound, there is no minima of energy. The meta-stable states appear in a region separating the whole phase space into two parts: stable and unbound.  In this plot we also exhibit the result obtained from a numerical solution of Eq. (\ref{eff})  with the  LHY interaction.  
A similar  phase plot based on  a numerical solution of  Eq. (\ref{present}) with  non-perturbative BMF interaction (\ref{ASmu})
 is shown in Fig. \ref{fig3}(b) for $\delta =4a_0$. For very  small values of the gas parameter $x=na^{1/3}$, this phase plot agrees with that in Fig. \ref{fig3}(a). For slightly larger values of the gas parameter, we have seen that the  $x^\nu, \nu=2.5$ divergence  of the universal function $f(x)$ (\ref{ASmu}), responsible for stopping collapse, disappears from the  BMF interaction (\ref{ASmu}), viz.  Fig. \ref{fig1}(b), and this happens for $x > 0.1$. Once this happens, the system collapses. The  collapse region is    illustrated in Fig. \ref{fig3}(b) for the BMF interaction (\ref{ASmu}), which is clearly absent in the case of the perturbative LHY interaction in Fig. \ref{fig3}(a).   For a fixed $\delta$, the collapse  takes place  for small $a$  
 due to a reduced repulsion. The self-bound state shrinks to a small size with higher density $n$ thus pushing $x=an^{1/3}$
 beyond 0.1 and causing the 
self-bound state to collapse. 
 The collapse instability is expected to be enhanced as the scattering length imbalance $\delta$ increases, consequently increasing the attraction and pushing the gas parameter $x$ beyond $x=0.1$. Once this happens the  non-perturbative BMF  interaction (\ref{present})  will not stop the collapse. Thus with added attraction   the domain of collapse has
increased in Fig. \ref{fig3}(c) for $\delta=8a_0$, compared to      Fig. \ref{fig3}(b) with
$\delta=4   a_0$. The
 unbound domain has also reduced in Fig. \ref{fig3}(c) with added attraction. With further increase of $\delta$ the collapse region increases.  

In Fig. \ref{fig4} we plot (a) root mean square radius $\langle r \rangle$ and (b) energy 
$E/N$  of the self-bound states for $\delta =4a_0$ and $N=100000$ for values of $a$ where  a stable state can be formed in the non-perturbative BMF model (\ref{present}).  We find that results for the 
non-perturbative and perturbative models qualitatively agree where a self-bound state can be formed. The energy of the crossover model in Fig. (\ref{fig4})(b) rapidly 
decreases as the scattering length $a$ is reduced signaling a collapse. { In the experiment of 
Semeghini et al.  \cite{inguscio} the value of $\delta$ was kept very small   and it was found that with the increase 
of $\delta$ the size of the binary self-bound state was drastically reduced before it was quickly destroyed
\cite{fattori}.  However, they did not study the fate of the state. Nevertheless, this behavior signals a collapse.  The 
three-body loss is a steady slow process, which will lead to a slower destruction of the self-bound state.  The value of $\delta$ was also kept small around $3a_0$ in the experiment of Cabrera et al.  \cite{Cabrera}. }

\section{Conclusion} We studied the formation of a self-bound state in a binary BEC using a realistic non-perturbative BMF interaction (\ref{ASmu})
and critically compared  the results 
with those obtained using the 
 perturbative LHY interaction (\ref{mu}). We find that  a self-bound state can be formed only for weakly attractive systems, where both  interactions could stop the collapse. For stronger attraction, the unrealistic LHY interaction continues to stop the collapse, whereas the realistic BMF interaction could not stop collapse and create a self-bound state. For an analysis of the self-bound BEC, a realistic non-perturbative  BMF interaction should be used in place of the perturbative  LHY interaction, which could lead to an inappropriate description. 

The present derivation of the BMF interaction (\ref{ASmu}) is heuristic, rather than rigorous,  in nature and has a single parameter $\eta$ fitted to the plausible   result of a microscopic calculation by Ding and Greene \cite{DG}.  Also, to keep the underlying model simple, in this application we considered a special case of a binary BEC with equal number of atoms $N_i$ and equal intra-species scattering lengths $a_i$ in the two components. 
Nevertheless, the disappearance of the LHY repulsion for values of the gas parameter $x\gtrapprox
0.01 $ and  the inability of the BMF  interaction to stop the collapse and support a self-bound state are  independent of the present derivation and subsequent calculation.  Similar conclusions demonstrating the  deficiency  of the LHY interaction to explain the properties of 
a self-bound state unconditionally   in a binary \cite{rec}   and in a dipolar  BEC \cite{pfau}   are in agreement with the conclusions of the present paper.  
 Our finding  is in agreement with the recent comparison of the LHY approximation, Monte Carlo 
simulations, and experiments on self-bound droplets in a dipolar gas \cite{pfau}
  demonstrating that the Monte Carlo simulations agree well with 
experiments, while the LHY approximation does not provide good agreement, being unable
to reproduce the observable properties of the quantum droplets. 
  Hence in spite of the heuristic derivation of our model and the special binary mixture considered here, we do not believe our conclusions to be so peculiar as to have no general validity.  
 Although we illustrated our findings for a binary BEC mixture, the conclusions will be applicable in general, for example, in the formation of a self-bound state in a dipolar BEC
\cite{Kadau} or in a binary boson-fermion  mixture \cite{arxiv}.

 \section*{Acknowledgements}
\noindent

%\begin{acknowledgements}
S.G. acknowledges the support of the Science \& Engineering Research Board (SERB), 
Department of Science and Technology, Government of India under the project 
ECR/2017/001436 and ISIRD project 9-256/2016/IITRPR/823 of Indian Institute of Technology (IIT) Ropar. 
S.K.A acknowledges the support by the Funda\c c\~ao de Amparo \`a Pesquisa do Estado de
S\~ao Paulo (Brazil) under project 2016/01343-7 and also by
the Conselho Nacional de Desenvolvimento Cient\'ifico e Tecnol\'ogico (Brazil) under project 303280/2014-0.
%\end{acknowledgements}

\end{document}